\documentclass{article}
\usepackage{ijcai17}
\usepackage{amsmath}
\usepackage{color}
\usepackage{times}
\usepackage{helvet}
\usepackage{courier}
\usepackage{booktabs}
\usepackage{epsfig}
\usepackage{amssymb}
\usepackage{amsmath}
\usepackage{amsfonts}
\usepackage{multirow}
\usepackage{array}
\usepackage{mathtools}
\usepackage{url}
\usepackage{epstopdf}
\usepackage{float}
\usepackage[]{algorithm2e}
\usepackage{flushend}
\usepackage[small]{caption}

\newcommand{\model}{SPMC} 
\newcommand{\eq}[1]{Eq.~(\ref{#1})}
\newcolumntype{R}[1]{>{\raggedleft\arraybackslash}p{#1}}
\newcommand{\xhdr}[1]{\vspace{2mm}\noindent{{\bf #1}}}
\DeclareMathOperator*{\argmax}{arg\,max}

\title{\model: Socially-Aware Personalized Markov Chains for Sparse Sequential Recommendation}
\author{Chenwei Cai, Ruining He, Julian McAuley \\ 
University of California, San Diego \\
\{cwcai, r4he, jmcauley\}@ucsd.edu}

\begin{document}

\maketitle
\begin{abstract}
%In recommender systems, people want to feed the systems with more information to improve the accuracy of the recommendations,  especially in \emph{cold-start} settings. 
Dealing with sparse, long-tailed datasets, and \emph{cold-start} problems is always a challenge for recommender systems.
%In current recommender systems, there are both sequential recommender systems, where feedback history is used, and social recommender systems, where social relations of the users are used. 
These issues can partly be dealt with by making predictions not in isolation, but by leveraging information from related events; such information could include signals from \emph{social relationships} or from the \emph{sequence} of recent activities.
%In both situations, additional information 
Both types of additional information can be used to
improve the performance of state-of-the-art matrix factorization-based techniques.
In this paper, we propose new methods 
to combine both social and sequential information simultaneously, in order to further improve recommendation performance.
%which take advantage of both sequential and social information in concert, %from datasets to further enhance the performance of the recommender system. 
%The feedback history of a user and that of the user's friends are both used in the algorithm. 
%The new recommender system should be able to handle user \emph{cold-start} issues in sparse datasets and real-world datasets.  
We show these techniques to be particularly effective when dealing with sparsity and \emph{cold-start} issues in several large, real-world datasets.

% We account for both fine-grained and high-level characteristics with a novel hierarchical embedding architecture. Our method outperforms state-of-the-art visually-aware models for the task of personalized ranking from real-world implicit feedback datasets.
\end{abstract}

\section{Introduction}
%1: High-level overview
\emph{Cold-start} 
%is always a barrier for recommender systems to obtain better performance. 
problems are a barrier to the performance of recommender systems that depend on learning high-dimensional user and item representations from historical feedback.
%JULIAN: Sentence doesn't say anything
%Great efforts have been made to pass through this barrier:\cite{}. 
In the one-class collaborative filtering setting, two 
%sorts of recommender systems are mostly used: 
types of techniques are commonly used to deal with such issues:
\emph{sequential} recommender systems and \emph{social} recommender systems. 
%The sequential recommender systems are modeled from feedback history, 
The former assume that users' actions are similar to those they performed recently, while the latter assume that users' actions can be predicted from those of their friends. In both cases, these related activities act as a form of `regularization,' improving predictions when users and items have too few observations to model them in isolation.
%The former model activities by making use of 
%JULIAN: Not sure what you're trying to say here
%where the sequences of the feedback are the key of the model. The sequantial recommender systems assume that users have similar preferences sequences. The social recommender systems, on the other hand, take in consideration the interactions between users to do predictions. 
%JULIAN: Doesn't say anything
%Both techniques are widely researched in \cite{}.

% The basic MF uncovers a set of \emph{latent} dimensions based on which an item or user is described by a latent feature vector, encoding the `properties' or `preferences' on each of these dimensions.

%2: What's the problem?
%In the techniques above, more parameters are introduced 
%JULIAN: Doesn't say anything beyond what's in the previous paragraph
% Models that include social and sequential regularization include additional parameters
% to improve 
% %the performance of the models. 
% model performance.
% More information is extracted from the datasets to deal with \emph{cold-start} issues which are brought about by the large number of parameters in the models. To further enhance the accuracy of (\emph{cold-start}) recommendation in sparse datasets, deeper extraction of the information and proper number of parameters are required in the new model.

%3: What's currently done? & What's its limitation?
Two existing models of interest are \emph{Factorized Personalized Markov Chains} (FPMC) \cite{rendle2010factorizing} and \emph{Social Bayesian Personalized Ranking} (SBPR) \cite{zhao2014leveraging}.
%In existing models, the most well-known sequential model is FPMC \cite{rendle2010factorizing}. 
In FPMC,
%In this work, 
the authors make use of `personalized' Markov chains
%. Both current basket as well as the last basket are used to train the model. 
to train on sequences of users' baskets;
FPMC is shown to enhance overall performance, 
%but this model is not good at handling \emph{cold-start} issues, 
though does little to address cold-start issues, due to the large number of parameters that need to be included to handle sequential recommendation.
%since too many latent factors need to be trained. There are more %attempts in social recommendation. 
Among a wide variety of work that makes use of social factors,
SBPR 
%in \cite{zhao2014leveraging} 
%assumes that the feedback on a item from a user's friends is more important than the item that neither viewed by the user nor the user's friends, but less important than the user's own feedback. 
uses a ranking formulation to assume that items considered by a user's friends are more important than items \emph{not} viewed by the user or their friends, but less important than items the user viewed themselves.
%Extending the state-of-the-art model BPR-MF, SBPR achieves higher accuracy in \emph{cold-start} settings. 
This is a natural extension of standard ranking-based objectives, such as Bayesian Personalized Ranking (BPR) \cite{rendle2009bpr}, and leads to increases in accuracy in cold-start settings. Similar techniques have also been used to rank and regularize based on other factors, such as groups of users who behave similarly \cite{pan2013gbpr}.
%Group recommendation is also studied in \cite{pan2013gbpr}, where the preference of a group of users is considered in the model.

%4: What our idea is?
\begin{figure}
\centering
\includegraphics[width=\columnwidth]{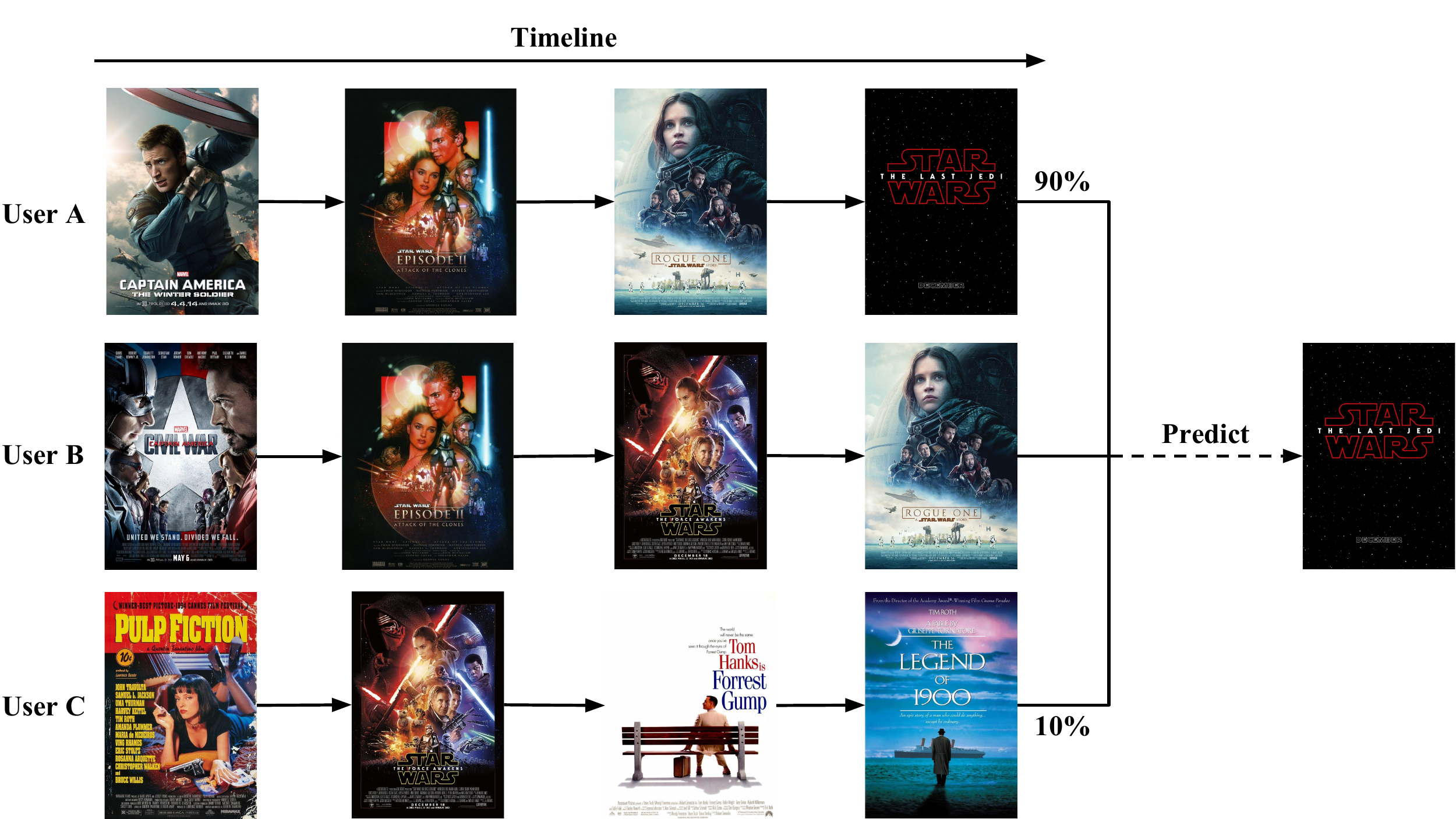}
%JULIAN: Figure doesn't strike me as very informative, beyond just saying that you use social+sequential. *How* is any of this achieved?
\caption{
% The high-level idea of \model~is illustrated above: our goal is to simultaneously model the sequential activities of a user and their friends. In the example above, both user A and user C are B's friends, but user A is much more similar to user B than user C; thus user A contributes more to user B's prediction. Furthermore, feedback sequences also contribute to the final prediction. In the example, the predictor compares the contribution from user A's sequence and user B's sequence, and makes a prediction on that basis.
High-level idea of the \model{} model. By simultaneously modeling the sequential activities of users and their friends, \model{} predicts the next action of a user (e.g.~B in this example) to be the combination of (1) her tastes (fantasy movies), (2) her own recent action (\textit{Rogue One}), and (2) those recent actions from her friends (i.e., \textit{The Last Jedi} from friend A, \textit{The Legend of 1900} from friend C).
\model{} learns to weigh different friends and make the final prediction (e.g.~A's recent activities are more influential on B in this example).}
\label{fig:illustration}
\end{figure}

\xhdr{In this paper} we propose a new model, \model~(Socially-aware Personalized Markov Chain) that leverages feedback from sequences, %but also utilizes the 
as well as
social interactions in the same model. The model is based on the assumption that a user can be affected both by their own feedback sequence as well as that of their friends' (Figure \ref{fig:illustration}). 
%We expect our new model can further 
Our goal in doing so is to improve upon
%improve the 
existing models, especially
when dealing with user \emph{cold-start} issues in sparse datasets. 
%JULIAN: Not really worth having a section like this in a very short paper.
% Our contributions include:
% \begin{itemize}
% \item We propose a new model, where the sequential and social information of data are both utilized. The number of parameters does not undermine the performance of the model on \emph{cold-start} recommendation.
% \item We evaluate the model on real-world datasets like Foursquare. Our model outperforms the existing models in \emph{cold-start} region.
% \end{itemize}

% \xhdr{Roadmap.} First, we describe related work in Section~\ref{sec:related_work}, before the \model~model is detailed in Section~\ref{sec:model}. Finally, we conduct comprehensive experiments in Section~\ref{sec:experiments}, where we compare \model~against some state-of-the-art models.

In essence, our model is a combination of FPMC and SBPR, such that users' actions are assumed to be determined by (1) their preferences; (2) their recent activities; and (3) their friends' recent activities. We make several simplifying assumptions to deal with the otherwise prohibitive number of parameters introduced by such a general formulation, especially when dealing with sparse datasets. Experiments on four real-world datasets reveal that the model is capable of beating state-of-the-art recommendation techniques that consider sequential and social factors in isolation.

\section{Related Work}
\label{sec:related_work}

The most closely related works to ours are (1) Item recommendation methods that model user preferences in terms of latent factors; (2) 
%Works that deal with temporal dynamics but rely on explicit time stamps; and 
Works that model sequential dynamics (and more broadly temporal information); and
(3) 
%Those that address the sequential prediction task we are interested in.
Socially-regularized recommender systems.

\xhdr{Item recommendation.} Item recommendation usually relies on Collaborative Filtering (CF) to learn from explicit feedback like star-ratings \cite{Handbook}. 
%CF predicts based only on the user-item rating matrix and mainly follows two paradigms: neighborhood- and model-based. 
%Neighborhood-based methods 
% recommended
%recommend
%Camera-ready: typo
%items that either have been enjoyed by like-minded users (user-oriented, e.g.~\cite{konstan1997grouplens,shardanand1995social,hill1995recommending}) or are similar to those already consumed (item-oriented, e.g.~\cite{AmazonItemToItem, deshpande2004item}). Such methods have to rely on some type of \emph{predefined} similarity metric such as {Pearson Correlation} or {Cosine Similarity}. 
%In contrast, 
Although several paradigms for explicit feedback exist, of most relevance to us are
\emph{model-based} methods,
%that directly explain the interactions between users and items. There have been a variety of such algorithms
including Bayesian methods \cite{miyahara2000collaborative,breese1998empirical}, Restricted Boltzmann Machines \cite{salakhutdinov2007restricted}, and in particular  Matrix Factorization (MF) methods (the basis of many state-of-the-art recommendation approaches such as \cite{BellKorSolution,Netflixprize,NSVD}).

Such models have been extended in order to tackle implicit feedback data where only positive signals (e.g.~purchases, clicks, thumbs-up) 
%are observed, both neighborhood- and model-based methods have been extended.
are observed (i.e., the so-called `one-class' recommendation setting).
%Recently, Ning \emph{et al.}~proposed SLIM to learn an item-item similarity matrix, which has shown to outperform a series of state-of-the-art recommendation approaches \cite{ning2011slim}. Kabbur \emph{et al.}~further explored the low-rank property of the similarity matrix to handle sparse datasets \cite{kabbur2013fism}. Since similarity (or neighborhood) relationships are learned from the data, these methods overcome the rigidity of using a predefined similarity metric. On the other hand, MF has also been extended in several ways including
Most relevant here are pair-wise methods like BPR-MF
%point-wise methods that inherently assume non-observed feedback to be negative \cite{WRMF,OCCF}, and pair-wise methods like BPR-MF 
\cite{rendle2009bpr} that 
%are based on a more realistic assumption that positive feedback should only be `more preferable' than non-observed feedback. 
make an assumption that positive feedback instances are simply `more preferable' than non-observed feedback. These are the same types of assumptions that have previously been adapted to handle implicit social signals, in systems like SBPR \cite{zhao2014leveraging}.

% \xhdr{Temporal dynamics.} 
% %Recent works
% Several works take temporal dynamics into account, mostly based on MF techniques \cite{koren2008factorization}. 
% %The most representative 
% This includes seminal
% work proposed by Koren~\cite{koren2010temporal,koren2009matrix}, where they showed state-of-the-art results on \emph{Netflix} data by modeling the evolution of users and items over time.
% %as time goes by. %colloquial
% However, such works are ultimately building models to understand 
% %the `past' 
% past actions
% (e.g.~`What did Tom like in 2008?', `What does Grace like to do on Weekends?'), by making use of the explicit time stamps. The sequential prediction task differs from theirs in that it does not use time stamps directly, but rather models sequential relationships between actions.
%and (2) It is more realistic as it aims to predict the `future'. %JULIAN I wouldn't say this is a real distinction

\xhdr{Sequential recommendation.} 
Markov chains are powerful methods for modeling stochastic transitions; they have been leveraged to model decision processes (e.g.~\cite{shani2002mdp}) and more generally uncover sequential patterns (e.g.~\cite{zimdars2001using,mobasher2002using}).
In the sequential recommendation domain, Rendle \emph{et al.}~proposed FPMC that combines MF and (factorized) Markov chains to be able to capture personalization and sequential patterns simultaneously \cite{rendle2010fpmc}. 
Our work follows this thread but extends these ideas by 
% (1) 
making use of social, in addition to sequential, dynamics. 
% and (2) we further consider Markov Chains with higher orders to 
% model sequential smoothness across multiple time steps.

\xhdr{Social recommendation.} %(maybe don't bother with this, but we should mention something related to modeling owners)
In the recommender systems literature, there has been a large body of work that models social networks for mitigating cold-start issues in recommender systems, e.g.~\cite{chaney2015probabilistic,zhao2014leveraging,ma2009learning,ma2008sorec,guo2015trustsvd}. 
% The type of social signals they usually benefit from are so-called `trust' relations amongst different users, much as we use here. 
For example, \emph{regularization}-based methods (e.g.~\cite{jamali2010matrix,ma2011recommender}) assume that users' preferences should be similar to those of their social circles. Given the social network information $\mathcal{F}_u$ of user $u$, this framework uses regularization to force $u$'s preference factors $\gamma^U_u$ to be close to those users in $\mathcal{F}_u$. Finally, feedback and social regularization are optimized simultaneously. Likewise, \emph{joint factorization}-based methods (e.g.~\cite{tang2013exploiting,ma2008sorec}) try to find a factorization of the social network matrix such that the resulting user representation can be directly used to explain users' preferences.

Our work differs from such socially-aware recommendation models mainly in that our method is sequentially-aware which not only makes it good at making predictions in a sequential manner---a desirable feature of recommender systems, but also perform well in cold-start settings. On the other hand, we propose to model the impact of the \emph{recent} activities of a user's friends on his/her own future activities, instead of assuming the closeness amongst social circles in terms of long-term preferences. 
In this paper, we also empirically compare against several state-of-the-art socially-aware recommendation methods and demonstrate the effectiveness of modeling such socio-temporal dynamics.

\begin{table}
\centering
\renewcommand{\tabcolsep}{2pt}
\caption{Notation \label{tab:notation}}
% \vspace{-3mm}
\begin{tabular}{lp{0.75\linewidth}} \toprule
Notation & Explanation\\ \midrule
$\mathcal{U}$, $\mathcal{I}$  & user set, item set\\
$u$, $i$  & user $u \in \mathcal{U}$; item $i \in \mathcal{I}$\\
$\mathcal{I}_u^+$ & `positive' item set for user $u$; $\mathcal{I}_u^+ \subseteq \mathcal{I}$\\
$\mathcal{F}_u$ & the friend set of user $u$;  $\mathcal{F}_u \subseteq \mathcal{U}$ \\
$\widehat{x}_{u,i,l}$ & predicted score $u$ gives to $i$ given last item $l$\\
$\beta_i$ & bias term associated with $i$; $\beta_i \in \mathbf{R}$ \\
$\gamma^U_u$, $\gamma^I_i$ & latent factors of $u$ / $i$; $\gamma^U_u,\gamma^I_i \in \mathbf{R}^{K_1}$\\
$\theta^I_i$, $\theta^L_i$ & latent factors of item $i$; $\theta^I_i,\theta^L_i \in \mathbf{R}^{K_2}$\\
$M_i$, $N_i$ & latent factors of item $i$; $M_i,N_i \in \mathbf{R}^{K_3}$\\
$W_u$, $V_u$ & latent factors of user $u$; $W_u,V_u \in \mathbf{R}^{K_4}$ \\ 
$K_1$, $K_2$, $K_3$& dimensionality of different latent factors\\
$\Theta$ & parameter set \\ 
$\alpha$ & hyperparameter weighting the influence of the friend set \\ 
$\sigma(\cdot)$ & sigmoid function; $\sigma(z) = 1/(1+e^{-z})$ \\
\textbf{1}$(\cdot)$ & indicator function; \textbf{1}$(b)$ = 1 \textit{iff} $b$ is $true$  \\ \bottomrule
\hline\end{tabular}
\end{table}

\section{The \model{}~Model} \label{sec:model}
In this paper, we focus on modeling implicit feedback, e.g.~clicks, purchases, or thumbs-up.
%, which are relatively abundant in the system.
In addition to the feedback itself, we assume that timestamps are also available for each action, as well as the social relations (or trust relationships) of each user.
%Timestamps associated with these feedbacks are also assumed to be available. In addition, the predictor is also fed with the social relations of each user. 

Let $\mathcal{U}$ denote the user set and $\mathcal{I}$ the item set.
%respectively. 
For each user $u\in\mathcal{U}$, we use $\mathcal{I}_u^+$ to denote the set of items toward which the user $u$ has expressed positive feedback. $\mathcal{F}_u$ is used to denote the set of users that user $u$ trusts, or the set of $u$'s friends. The objective of our task is to predict the sequential behavior of users given the above information on \emph{sparse} datasets, where dealing with user \emph{cold-start} issues is paramount.

Notation used throughout this paper is summarized in Table \ref{tab:notation}.

\begin{figure*}
\begin{equation} \label{eq:SPMC-1}
\widehat{x}_{u,i,l} = \underbrace{\langle\gamma^U_u, \gamma^I_i\rangle}_{\text{user preferences}} + \underbrace{\langle\theta^I_i, \theta^L_l\rangle}_{\text{sequential dynamics}} + ~\underbrace{\frac{2}{|\mathcal{F}_u|^\alpha}\sum_{u'\in \mathcal{F}_u, i'}\sigma\big(\langle W_u, V_{u'}\rangle \big) \cdot \langle M_i, N_{i'} \rangle}_{\text{socio-temporal dynamics}} ~+ \underbrace{\beta_i}_{\text{item bias}} 
\end{equation}
\caption{The proposed socially- and sequentially-aware predictor.}
\end{figure*}

% \subsection{Form of the Original \model{}~Model}
\subsection{Model Specifics}
Our \model{} model is built on top of a state-of-the-art sequential predictor named \emph{Factorized Personalized Markov Chains} (FPMC) \cite{rendle2010factorizing}. 
In FPMC, given a user $u$ and the last item they interacted with $l \in \mathcal{I}_u^+$, the probability that $u$ transitions to another item $i$ is proportional to 
\begin{equation} \label{eq:FPMC}
\widehat{x}_{u,i,l} = \langle\gamma^U_u, \gamma^I_i\rangle + \langle\theta^I_i, \theta^L_l\rangle, 
\end{equation}
where the first inner product models the `affinity' between latent user factors $\gamma^U_u\in\mathbf{R}^{K_1}$ and latent item factors $\gamma^I_i\in\mathbf{R}^{K_1}$, and the second models the `continuity'  between item $i$ and the previous item $l$. $\theta^I_i, \theta^L_l\in\mathbf{R}^{K_2}$ are latent representations of item $i$ and the last item $l$ respectively.

The above predictor is capable of capturing both personalization and sequential dynamics. However it is unaware of the social signals in the system which are potentially important side information especially in cold-start scenarios. In particular, we argue that the recent actions of a user's friends could be influential when 
%the user decides on moving forward to the next actions.
determining which action a user is likely to perform next.

Our \model{} model is a combination of personalization, sequential dynamics, as well as socio-temporal dynamics. The predictor of our model is simply the 
sum of three such components (see \eq{eq:SPMC-1}),
%addition of the three components:
% \begin{equation} \label{eq:SPMC-1}
% \begin{array}{rl}
% \widehat{x}_{u,i,l} = & \underbrace{\langle\gamma^U_u, \gamma^I_i\rangle}_{\text{user tastes}} + \underbrace{\langle\theta^I_i, \theta^L_l\rangle}_{\text{sequential dynamics}} + \\
% &\underbrace{\frac{2}{|F_u|^\alpha}\sum_{u'\in F_u, i'}\sigma(\langle W_u, V_{u'}\rangle) \cdot \langle M_i, N_{i'} \rangle}_{\text{socio-temporal dynamics}} + \beta_i, 
% \end{array}
% \end{equation}
% \begin{figure*}
% \begin{equation} \label{eq:SPMC-1}
% \widehat{x}_{u,i,l} = \underbrace{\langle\gamma^U_u, \gamma^I_i\rangle}_{\text{user tastes}} + \underbrace{\langle\theta^I_i, \theta^L_l\rangle}_{\text{sequential dynamics}} + \underbrace{\frac{2}{|F_u|^\alpha}\sum_{u'\in F_u, i'}\sigma(\langle W_u, V_{u'}\rangle) \cdot \langle M_i, N_{i'} \rangle}_{\text{socio-temporal dynamics}} + \beta_i, 
% \end{equation}
% \end{figure*}
where
% \begin{equation} \label{eq:SPMC-2}
% T_{u'}(i') = \max\{T_{u'}(i_0) | T_{u'}(i_0) < T_u(i) \wedge i_0 \in \mathcal{I}_{u'}^+\},
% \end{equation}
% and $T_u(i)$ is the timestamp of user $u$'s feedback on item $i$.
%CHENWEI 5/2/2017: changed from: $i'$ is the item viewed by one of $u$'s friends $u'$ most recent to user $u$'s feedback on item $i$.
$i'$ is the item viewed by one of $u$'s friends $u'$ most recent to user $u$'s feedback on item $i$.
%where $i'$ is 
%JULIAN: The word "immediately" doesn't seem to make sense here.
%immediately 
%JULIAN: Seems redundant.
%before $i$ on the timeline. 
The inner product of the latent factors of $i$ and $i'$ ($M_i,N_{i'}\in\mathbf{R}^{K_3}$) models the impact from friends' recent actions. 

Note that intuitively different friends could have a different amount of impact on a user. Therefore we measure the `closeness' between two users by the inner products of their latent representations ($W_u, V_{u'}\in\mathbf{R}^{K_4}$), which are normalized via a sigmoid function ($\sigma(\cdot)$) to be between 0 (no influence) and 1 (high influence).
%The 
%logistic 
%JULIAN: Not clear why this step is needed
%sigmoid function $\sigma(\cdot)$ is adopted to normalize the value to between 0 and 1, which helps with the convergence of the learning algorithm. 

Finally, a bias term $\beta_i\in\mathbf{R}$ is added to the formulation to capture the overall popularity of item $i$. $\alpha\in\mathbf{R}$ is a 
%JULIAN: Why is it a hyperparameter and not simply a parameter? Is it selected using the validation set?
%hyper
parameter that balances social effects against other factors.
%different terms.
%\underbrace{\langle \overbrace{\theta_u;\gamma_u}^{\text{user factors}}, \overbrace{\theta_i;\gamma_i}^{\text{item factors}} \rangle}_{\text{interactions between $u$ and $i$}} + \underbrace{\langle \vartheta, f_i \rangle}_{\text{visual bias of $i$}} + \beta_i,

\subsection{Merging the Embeddings}
Our goal is to deal with user \emph{cold-start} issues (i.e., users who have performed few previous actions), which requires us to reduce the number of parameters to be inferred. To this end, we merge the embeddings in \eq{eq:SPMC-1} and obtain the following predictor:
\begin{multline} \label{eq:SPMC-3}
% \begin{array}{rl}
\widehat{x}_{u,i,l} = \langle\gamma^U_u, \gamma^I_i\rangle + \langle\theta^I_i, \theta^I_l\rangle + \\
\frac{2}{|\mathcal{F}_u|^\alpha}\sum_{u'\in \mathcal{F}_u, i'}\sigma\big(\langle W_u, W_{u'}\rangle \big) \cdot \langle M_i, M_{i'} \rangle + \beta_i.
% \end{array}
\end{multline}
Note that the above equation merges (1) $\theta^I$ and $\theta^L$ into the same space $\theta^I$, (2) $W$ and $V$ into the same space $W$, and (3) $M$ and $N$ into the same space $M$. Further merges are possible but 
avoided here as they empirically resulted in degraded performance, presumably because they 
%hurt the expressive power too much. 
sacrifice too much of the model's expressive power.

% \subsection{Learning from Sequence and Social Relations}
% \model{} learns both from sequence and social relations. To learn a model for a user $u$, it first learns from the local feedback sequence as FPMC does. Then the model retrieves the last items of user $u$'s friend from $u$'s social relation list. The weights and the features of friends' last items are learned by the similar process as FPMC. Therefore, the social relation term can be regarded as an extension to local FPMC.

\subsection{Learning the Model}
An advantage of \model{} is that the same training framework from FPMC can be used to optimize the personalized total order $>_{u, l}$. In particular, %
%CHENWEI 5/2/2017: changed from "assuming that all users are independent and for each user all adjacent item pairs are independent"
%JULIAN: Don't really understand. What does it mean for users and items to be independent? And how is this not an assumption?
%all users and items are independent, so that 
Maximum a Posteriori (MAP) estimation of our parameters can be formulated as:
\begin{equation} \label{eq:MAP}
\begin{aligned}
\argmax_{\Theta} &= \ln \prod_{u \in \mathcal{U}} \prod_{i \in \mathcal{I}_u^+} \prod_{j \neq i} p(i >_{u, l} j | \Theta) ~ p(\Theta) \\
&=\sum_{u \in \mathcal{U}} \sum_{i \in \mathcal{I}_u^+} \sum_{j \neq i} \ln p(i >_{u, l} j | \Theta) + \ln p(\Theta),
\end{aligned}
\end{equation}
where $l$ is the item preceding $i$ in user $u$'s feedback sequence. The probability that user $u$ prefers item $i$ over item $j$ given $u$'s last item $l$ is 
%formulated by a sigmoid function:
given by
\begin{equation}
p(i >_{u, l} j | \Theta) = \sigma(x_{u,i,l} - x_{u,j,l}).
\end{equation}

The full set of parameters of \model{} is $\Theta = \{\beta_{i\in \mathcal{I}}, \gamma_{u\in \mathcal{U}}^U, \gamma_{i\in \mathcal{I}}^I, \theta_{i\in \mathcal{I}}^I, W_{u\in \mathcal{U}}, M_{i\in \mathcal{I}}\}$. We uniformly sample from the dataset a user $u$, a positive item $i$ and an item $j$ that is different from the positive item,
following the same negative-item selection protocol used in FPMC.
%. Item $j$ is the same as the selection of item $j$ in FPMC, where the item $j$ is not in the basket. 
%JULIAN: Shouldn't you select a negative item, rather than one that is different from the positive item? 
%CHENWEI: This is the same as the `negative' item in FPMC with only one item in the basket.
%about which the user has not expressed positive feedback.
% The last item $l$ can be obtained from the user's timestamped feedback history. The user's friends $F_u$ are retrieved from the social relation list of the user.
Since the last item $l$ and the relevant actions from the friends $\mathcal{F}_u$ are determined by the pair $(u, i)$, we can apply standard stochastic gradient ascent to optimize \eq{eq:MAP} and update the parameters as
\begin{equation}
\Theta^{t+1} \leftarrow \Theta^t + \eta \cdot \big(\sigma(\widehat{x}_{u,j,l} - \widehat{x}_{u,i,l}) \frac{\partial (\widehat{x}_{u,i,l} - \widehat{x}_{u,j,l})}{\partial \Theta} - \lambda_{\Theta}\Theta \big), 
\label{eq:update}
\end{equation}
where $\eta$ is the learning rate and $\lambda_{\Theta}$ is a regularization hyperparameter. For completeness, we list the partial derivative of $\widehat{x}_{u,i,l} - \widehat{x}_{u,j,l}$ with respect to our parameters in Appendix A.
%term to prevent overfitting.

\section{Experiments}
\label{sec:experiments}

To fully evaluate the effectiveness of our proposed model, we perform extensive experiments on a series of real-world datasets and compare against state-of-the-art sequentially- and socially-aware methods. 

\subsection{Datasets}
We experiment on four datasets, each comprising a large corpus of user feedback, timestamps, as well as social relations (i.e.~`trusts'). All datasets are available online.

\xhdr{Ciao.} Ciao is a review website where users give ratings and opinions on various products. This dataset was crawled by \cite{tang2012mtrust} from Ciao's official site.\footnote{\url{http://www.ciao.co.uk/}} The feedback was given in the month of May, 2011. The dataset is available online.\footnote{\url{http://www.cse.msu.edu/~tangjili/trust.html}}

\xhdr{Foursquare.} This dataset is from Foursquare.com\footnote{\url{https://foursquare.com/}} and consists of check-ins of users at different venues, spanning December 2011 to April 2012. 
%Note that the setting what we are focusing on is to use collaborative data and social signals exclusively---taking advantage of other signals like geographical data is not within the scope of this paper.
While the dataset includes features other than just social and sequential signals (such as geographical data), these are beyond the scope of this paper.

\xhdr{Epinions.} Epinions is a popular online consumer review website. Collected by \cite{zhao2014leveraging}, this dataset also contains trust relationships amongst users and spans more than a decade, from January 2001 to November 2013.

\xhdr{Flixster.} Flixster is a social movie website where users can rate movies and share their reviews. This dataset is available online.\footnote{\url{http://www.cs.ubc.ca/~jamalim/datasets/}}
% CHENWEI: http://www.cs.ubc.ca/~jamalim/datasets/ this link is not available now. I cannot check for more details about this dataset.

In our experiments, we 
%take all ratings as `positive' feedback, regardless of the specific values, since users have made decisions to interact with those items after all, e.g.~purchases, watches. 
treat all observed interactions (i.e., ratings etc.) as positive instances, such that our goal is to rank items that a user would be likely to interact with.
%(as opposed to, say, predicting ratings).

% As we focus on on sparse datasets, we fed Foursquare to the model. The data set contains (1) users' feedback on items, with timestamps on each feedback (2) social relations of users, represented by a directed graph. We excluded all user-item pairs whose ratings are less than 4 out of 5, because these records cannot reflect users' preferences on these items. We also processed the data set to remove the users who only have less than 4 items in their feedback history. As for the evaluation protocol described in next part, the data set can have a sequence for every user only after this step.

\begin{table}
\centering
\renewcommand{\tabcolsep}{1.4mm}
\caption{Dataset statistics} 
\begin{tabular}{lrrrrr} \toprule
 & Ciao & Foursquare & Epnions & Flixster\ \\\midrule
\#users & 1,708 & 48,976 & 5,261 & 60,692\\
\#items & 16,485 & 29,879 & 25,996 & 48,549\\
\#feedback & 34,494 & 262,105 & 46,732 & 7,956,727\\
\#trusts & 35,753 & 231,506 & 23,915 & 840,141\\
\#votes/\#users & 20.20 & 5.35 & 8.88 & 131.10\\
\#trusts/\#users & 20.93 & 4.72 & 4.55 & 13.84\\
\bottomrule
%             &\#users ($|\mathcal{U}|$) & \#items ($|\mathcal{I}|$)   &\#feedback &\#trusts  \\\midrule
%Foursquare       & 48,976  & 29,879   & 262,105 & 231,506  \\\bottomrule
 \end{tabular}
\label{table:dataset}
\end{table}

\subsection{Evaluation Protocol}
The evaluation protocol we adopt is similar to previous works on sequential predictions, e.g.~\cite{vista}. From each user one positive item $\mathcal{T}_u\in\mathcal{I}_u^+$ is held out for testing, and another item $\mathcal{V}_u\in\mathcal{I}_u^+$ is held out for validation. The test item $\mathcal{T}_u$ is chosen to be the most recent item according to user $u$'s feedback history, while the validation item $\mathcal{V}_u$ is the second most recent one. The rest of the data is used as the training set. We report the accuracies of all models in terms of the AUC (\textit{Area Under the ROC Curve}) metric on the test set:
\begin{equation}
\mathit{AUC} =  \frac{1}{|\mathcal{U}|}  \sum_{u\in\mathcal{U}}   \frac{1}{|\mathcal{I}\setminus\mathcal{I}_u^+|}   \sum_{j \in \mathcal{I} \setminus \mathcal{I}_u^+}  \mathbf{1} (\widehat{x}_{u,\mathcal{T}_u,l_{\mathcal{T}_u}} > \widehat{x}_{u,j,l_{\mathcal{T}_u}}),
\end{equation}
where $\mathbf{1}(\cdot)$ is the indicator function, and $l_{\mathcal{T}_u}$ is the item preceding the test item $\mathcal{T}_u$ in $u$'s feedback history.

%Recall at $K$ ($R@K$) are used as an additional metric for performance evaluation. 

\subsection{Baselines}
We compare against state-of-the-art recommendation models, i.e.~BPR-MF and FPMC, as well as the socially-aware methods including GBPR and SBPR. 
\begin{itemize}
\item \textbf{BPR-MF}: This model is described in \cite{rendle2009bpr}, which is a state-of-the-art matrix factorization method that focuses on modeling user preferences.
\item \textbf{FPMC}: This model is described in \cite{rendle2010factorizing}, which is a state-of-the-art sequential prediction model. It unifies the power of matrix factorization at modeling users' preferences and the strength of Markov chains at capturing sequential continuity. Its predictor is presented in \eq{eq:FPMC}.
\item \textbf{SBPR}: This is a state-of-the-art recommendation model that benefits from modeling social relations. Introduced by \cite{zhao2014leveraging}, the model is based on the assumption that users and their social circles should have similar tastes/preferences towards items.
\item \textbf{GBPR}: A well-known model introduced by \cite{pan2013gbpr}. It makes use of group information, where users in a group have positive feedback on the same item.
\end{itemize}

Note that all methods 
%are directly 
optimize the AUC metric on the training set and the best hyperparameters are selected with grid search using the validation set.

%The design of the 
The
above baselines are selected to demonstrate that \model{} can outperform (1) state-of-the-art matrix factorization models that are unaware of sequential and social signals (i.e., BPR-MF); (2) methods that model both user preferences and sequential dynamics but no social signals (i.e., FPMC); and (3) models that are aware of social signals but ignore sequential dynamics (i.e.,~SBPR and GBPR).

\subsection{Performance Analysis}
\begin{figure*}
\centering
\includegraphics[width=\textwidth]{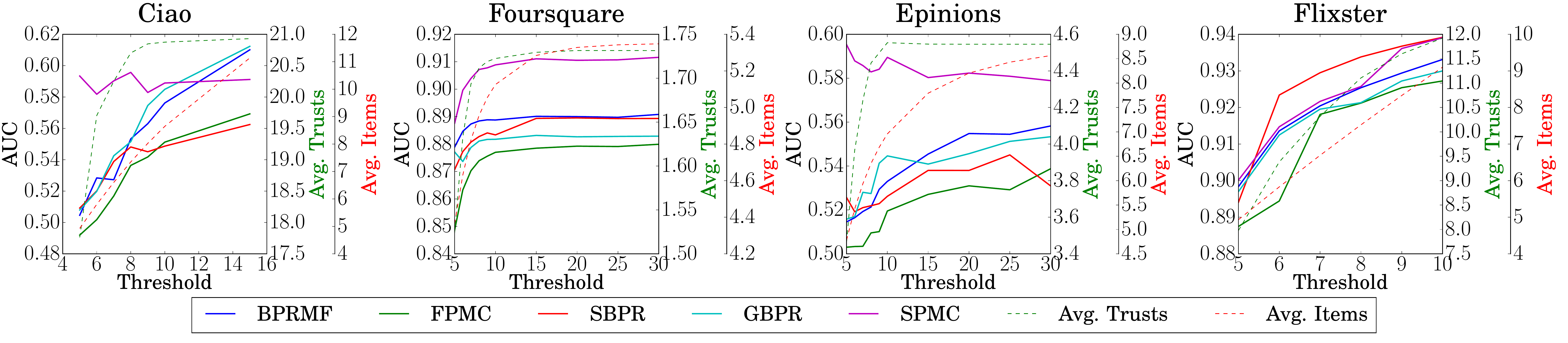}
\caption{
%The average AUC on 4 different datasets. 
Average AUCs on Ciao, Foursquare, Epinions and Flixster as dataset sparsity varies. 
% (other datasets yielded similar results).
%JULIAN: This figure isn't showing enough to warrant a full page.
% The models used to evaluate as shown in the legend. 
The $x$-axis is the threshold $N$ controlling the amount of feedback used. Each dataset is filtered according to $N$, resulting in datasets where the amount of feedback per user does not exceed the threshold. 
%JULIAN: Too repetitive
%In every dataset, we compare \model{} with the baselines in different threshold settings. 
% Since we are focusing on user \emph{cold-start} issues, we select the range of threshold where the data sets are sparse. 
Dashed green lines show the average number of social relations for the users, while dashed red lines show the average amount of feedback per user.}
\label{fig:auc}
\end{figure*}

% We train the models based on their overall performance according to average AUC. The AUC of the cold users from each models are then extracted. 
Our goal is to mitigate the user cold-start problem which is present in many real-world datasets. To this end, for each of the four datasets introduced earlier, we obtain a series of user cold-start datasets by varying a threshold $N$. 
% We filter the datasets with different thresholds to keep the users $cold$. 
$N$ is the maximum number of recent feedback instances each user can keep in his/her feedback history. In other words, if the number of user $u$'s observed interactions exceeds the threshold $N$ in the original dataset, only the most recent $N$ instances will be kept.\footnote{If the number of $u$'s feedback instances is less than the threshold $N$, all the feedback will be used.} %
%CHENWEI 5/2/2017: Added
The threshold $N$ is no less than 4 since we need at least 4 feedback instances for a user to form a sequence in the training set.
% Therefore, after this kind of filtering with threshold $N$, the average items that each user has reviewed is less or equal than $N$.
This protocol allows us to measure the performance improvements as the level of `coolness' varies.

For fair comparison, we set the dimensions of latent factors in all models to 20, i.e.,~for \model{} $K_1=K_2=K_3=20$. 
% Setting the dimensions to 20 can avoid lack of expressive power and the curse of dimensionality.
Empirically, using larger dimensionality did not yield 
significant improvements for any of the methods being compared, presumably because these datasets are sparse and 
%too many parameters are unaffordable.
a large number of parameters would be unaffordable.
%As for the learning rate, we tried 
We experimented with learning rates
$\eta \in \{0.5, 0.05, 0.005\}$ 
%and chose the one that leads to the best accuracy. Similarly, we also selected regularization hyperparameter term 
and regularization hyperparameters
$\lambda \in \{1, 0.1, 0.01, 0.001\}$, selecting the values that resulted in the best performance on the validation set. We set $\alpha=1$ for all datasets (the sensitivity of $\alpha$ will be discussed later). For GBPR, the group size is set to 3 and $\rho$ is set to 0.8. We implemented SBPR-2 as described in the original paper \cite{zhao2014leveraging}.

Table~\ref{table:auc} shows the average AUCs of all models on the test sets as we vary the threshold $N$. We give the percentage improvement of our model over FPMC as well as the best performing baselines in the last two columns of this table. Figure~\ref{fig:auc} shows the trends of average AUCs with the thresholds. From these results, we find that:
\begin{itemize}
\item As shown in Table~\ref{table:auc}, in very cold settings where thresholds are set to 5, \model{} always outperforms baselines.
\item Figure~\ref{fig:auc} and Table~\ref{table:auc} both show that \model{} can outperform other models in cold-start regions (e.g.~threshold from 5 to 10) on most datasets.
\item Figure~\ref{fig:auc} also shows that \model{} can significantly outperform other baselines on Foursquare and Epinions even when the threshold is set to large values. 
% Both the average number of user social relations and the average number of user feedback are abundant enough to support \model{}.
By comparing Foursquare and Epinions with the other datasets, we can see that the  average number of items per user is comparatively much lower on these two datasets. This means that the majority of users on Foursquare and Epinions are actually `cold' and thus they favor models that are strong in handling such cases.
\item \model{} can always outperform state-of-the-art socially-unaware sequential method---FPMC. This means that it is important to model social signals in order to benefit from such auxiliary information especially in cold-start settings.
% \item In very cold regions, e.g. threshold equals 5, BPR-MF has higher AUC than FPMC, since BPR-MF has less parameters to be trained. On the other hand, in very cold region, the social recommender systems can outperform the state-of-the-art methods in most of cases, because social information is used. Our model is always the best among the others in very cold regions. This can be explained by the fact that our model mixes the sequential information with the social information from the datasets.
\end{itemize}
% As we expected, \model{} outperforms the baselines in user \emph{cold-start} settings in real-world datasets.

In conclusion, by combining personalization, sequential dynamics, and socio-temporal information carefully, our proposed model \model{} considerably outperforms all baselines that model these signals in isolation in user \emph{cold-start} settings.

\begin{table*}
\renewcommand{\tabcolsep}{6pt}
\centering
\caption{AUCs of all models on the test sets as the threshold $N$ varies. The last two columns compare \model{} to FPMC as well as the best performing baseline. The best performing method is boldfaced (higher is better).}
\begin{tabular}{lcccccccccccccccccc} \toprule
\multirow{2}{*}{Dataset} &\multirow{2}{*}{Threshold ($N$)}   &(a)    &(b)    &(c)     &(d)    &(e)   &\multicolumn{2}{c}{improvement} \\ 
           &          &BPR-MF   &FPMC  &SBPR   &GBPR  &\model & e vs. b  & e vs. best \\ \midrule
\multirow{3}{*}{Ciao} & 5 &0.504614 &0.491940 &0.509185  &0.507940 &\textbf{0.593383}  &20.62\%  &16.54\%  \\
					  & 10 &0.576186 &0.551231 &0.548679  &0.584891 &\textbf{0.588934}  &6.84\%  &0.69\%  \\
                      & 15 &0.610123 &0.569248 &0.562466  &\textbf{0.612324} &0.591179  &3.85\%  &-3.45\%  \\
[4pt]
\multirow{3}{*}{Foursquare} & 5 &0.878899 &0.848644 &0.870693  &0.877121 &\textbf{0.887596}  &4.59\%  &0.99\%  \\
					        & 10 &0.888771 &0.876997 &0.883340  &0.881735 &\textbf{0.908859}  &3.63\%  &2.26\%  \\
                            & 15 &0.890104 &0.878499 &0.889350  &0.883149 &\textbf{0.911066}  &3.71\%  &2.36\%  \\
[4pt]
\multirow{3}{*}{Epnions} & 5 &0.514541 &0.503011 &0.525640  &0.515770 &\textbf{0.595317}  &18.35\%  &13.26\%  \\
						 & 10 &0.532994 &0.519503 &0.526204  &0.544611 &\textbf{0.589500}  &13.47\%  &8.24\%  \\
                         & 15 &0.545447 &0.527076 &0.538011  &0.540879 &\textbf{0.580287}  &10.10\%  &6.39\%  \\
[4pt]
\multirow{2}{*}{Flixster} & 5  &0.898370 &0.887483 &0.894095  &0.897011 &\textbf{0.900146}  &1.43\%  &0.20\%  \\
					     & 10  &0.929493 &0.927250 &\textbf{0.939176}  &0.929989 &0.939046  &1.27\%  &-0.01\%  \\\bottomrule
\end{tabular}
\label{table:auc}
\end{table*}

%JULIAN: Not very important, doesn't show anything other than what we'd expect in a typical learning setting
% \subsection{Training Efficiency}
% Figure~\ref{fig:curve} shows the learning curve of \model{} on the datasets when the thresholds are set to 5. In user \emph{cold-start} setting, our method converges with in 100 iterations on all datasets except for Foursquare, where a larger number of iterations spent because of the low learning rate. The fact indicates that the training efficiency and the convergence of the model is not largely impacted by the selection of data sets.
% \begin{figure}[H]
% \centering
% \includegraphics[width=.495\columnwidth]{curves/Ciao_curve.pdf}
% \includegraphics[width=.495\columnwidth]{curves/Foursquare_curve.pdf}\\
% \includegraphics[width=.495\columnwidth]{curves/Epinions_curve.pdf}
% \includegraphics[width=.495\columnwidth]{curves/Flixster_curve.pdf}
% \caption{The learning curve of \model{} on different datasets when the thresholds are 5.}
% \label{fig:curve}
% \end{figure}
\subsection{Convergence}
\begin{figure}
\centering
\includegraphics[width=\columnwidth]{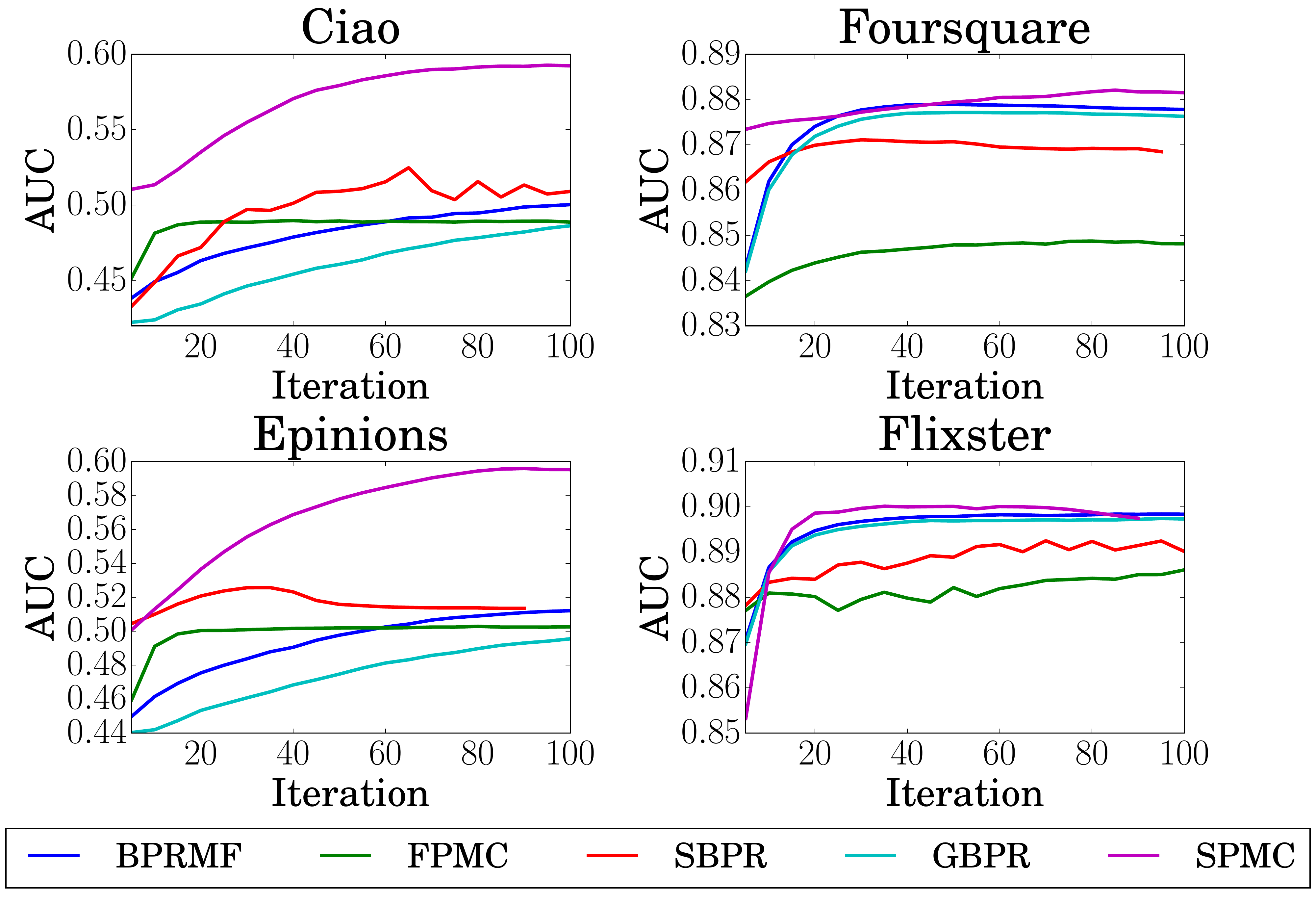}
\caption{Learning curves of \model{} and baselines on Ciao, Foursquare, Epinions and Flixster. The threshold $N$ is set to 5. \model{} converges roughly as fast as other methods.}
\label{fig:curve}
\end{figure}
% In \model{}, for each iteration, one quadruples $(u, i, j, l)$ is sampled from the whole dataset corpus, and then extract user $u$'s social relations from the dataset. This process is comparable to sampling a triplet $(u, i, j)$ from corpus in BPR-MF. Therefore, \model{} should converge at the same pace as the baselines do. 
We proceed by demonstrating comparison of convergence rates of all methods on the four datasets. Figure~\ref{fig:curve} shows the AUCs of \model{} and baselines on the test sets when the thresholds are set to 5. 
As we can see from this figure, the convergence efficiency of \model{} is comparable to all baselines (converging in fewer than 100 iterations),\footnote{Note that each training iteration (of all methods) is a sweep of all positive feedback in the training set.} although it is the integration of multiple sources of signals and is comparably complicated in its form.

% \subsection{Impact of $\alpha$ on Performance of \model{}}
% In equation~\ref{eq:SPMC-1}, the averaged behavior of user's friends is added as one term of the total score by the number of user $u$'s friends $|F_u|$. We also introduce one hyperparameter $\alpha$ to further tune contribution of the score from user $u$'s friends. Figure~\ref{fig:alpha} shows the impact on AUC when $\alpha$ varies. We select $\alpha\in\{0.6, 0.8, 1, 1.2, 1.4, 1.6\}$ in the model.
% \begin{figure}
% \centering
% \includegraphics[width=\columnwidth]{img/alpha.pdf}
% \caption{The impact of $\alpha$ on the AUC of \model{} on four datasets. Different selections of $\alpha{}$ have slight impact on the AUC.}
% \label{fig:alpha}
% \end{figure}

\begin{figure}
\centering
\includegraphics[width=\columnwidth]{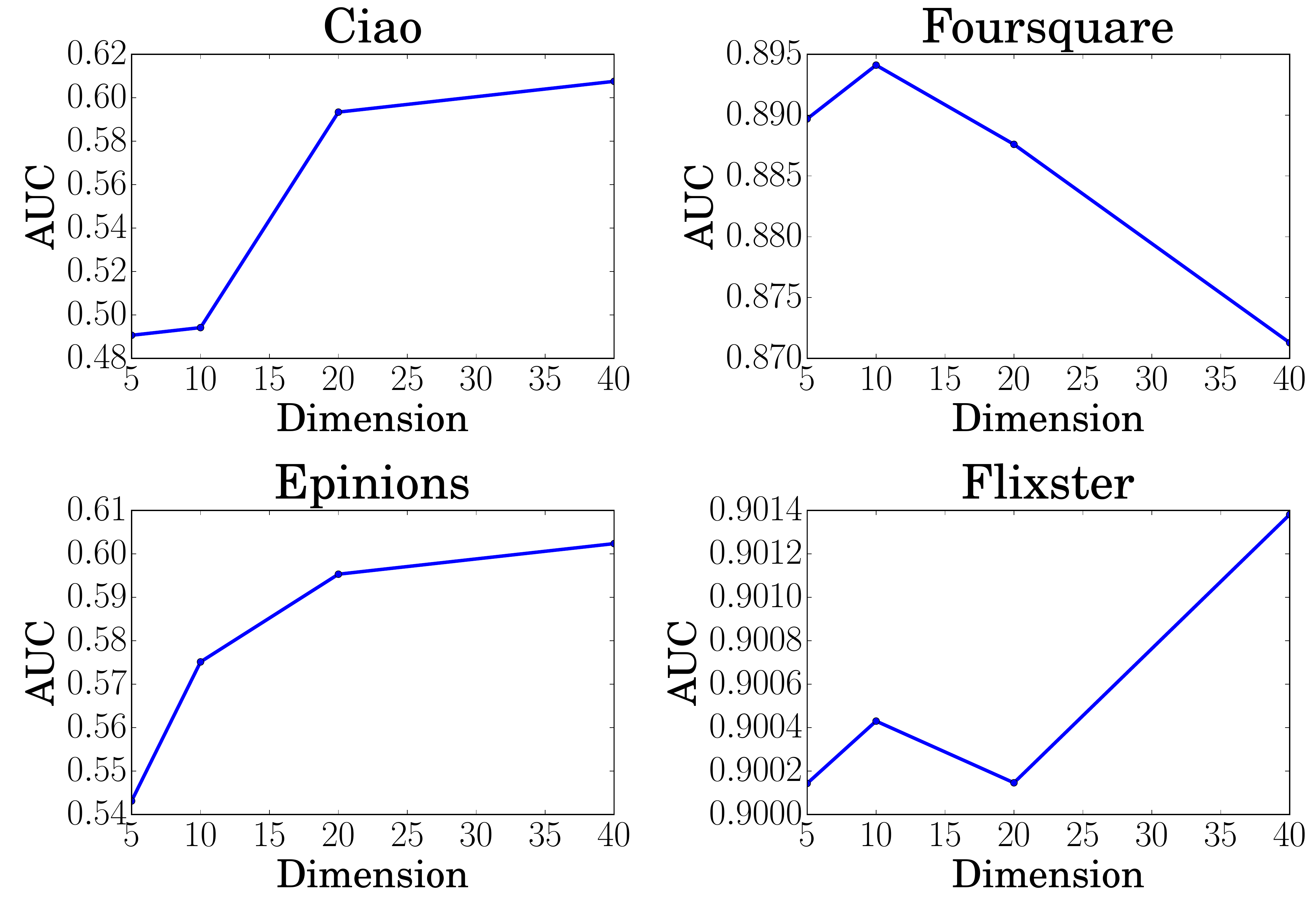}
\caption{The AUCs of \model{} on four datasets when the dimensionality $K$ varies. The model is more stable on larger datasets---Foursquare and Flixster. Note that AUCs are changing on a smaller scale on Foursquare and Flixster.}
\label{fig:hyper-k}
\end{figure}

\subsection{Sensitivity}
%\subsubsection{a}
Next we demonstrate the sensitivity of \model{} to different hyperparameters. Figure \ref{fig:hyper-k} shows the changes in AUC of \model{} on the four datasets as the dimensionality (here we adopt $K_1 = K_2 = K_3 = K$) increases from 5 to 40. We keep $\alpha = 1$, threshold $N=5$, and set $K \in \{5, 10, 20, 40\}$.
%where $K$ grows exponentially. 
From the figure we can see that the AUC does not improve significantly when the number of dimensions is larger than 20 in most cases, which is expected as each user is only associated with around 5 activities. In addition, it seems that our model is more stable on Foursquare and Flixster, presumably because their sizes are much larger than Ciao and Epinions.
% The result indicates that our model is quite stable on large datasets like Foursquare and Flixster, but less stable on the small ones like Ciao and Epinions.  

\begin{figure}
\centering
 \includegraphics[width=\columnwidth]{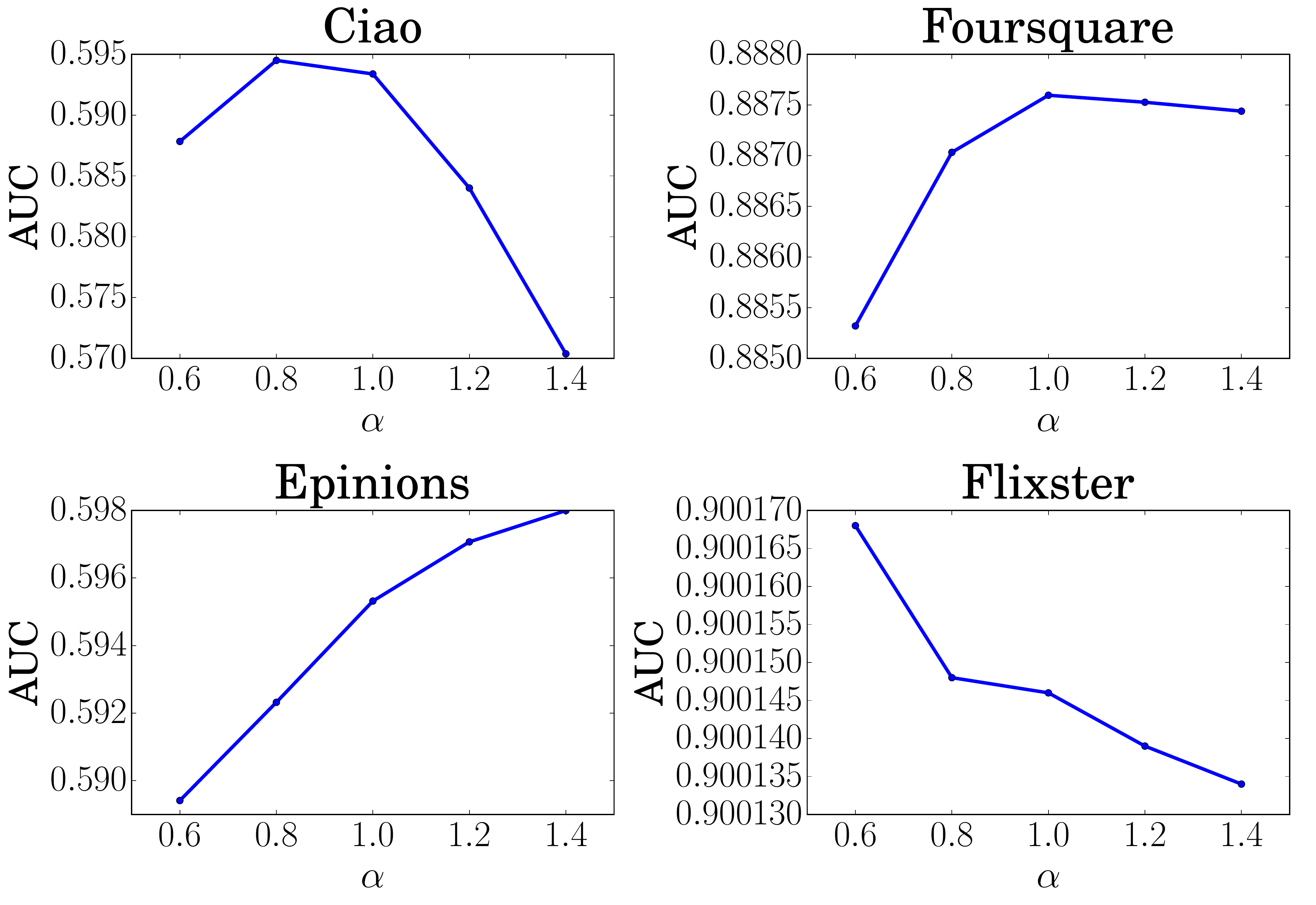}
\caption{AUCs on the four datasets achieved by \model{} as we vary the hyperparameter $\alpha$. Note that $\alpha$ seems to be the best when it is around 1.0, which essentially means that we are simply averaging the impacts from all friends.}
\label{fig:hyper-a}
\vspace{-2mm}
\end{figure}

We show the variation in AUCs of \model{} as we vary the values of $\alpha$ in Figure \ref{fig:hyper-a}. Empirically, it seems that the best performance can be achieved when $\alpha$ is around 1.0. This makes sense as it essentially means that we should take the Arithmetic Mean of the impacts from each friend.

%\textcolor{red}{Currently tweaking the graphs for tuning the hyperparameters $\alpha$ and $K$ ...}

\section{Conclusion}
In this paper, we proposed a new method, \model, to exploit both sequential and social information 
%from the datasets. 
for recommendation.
By combining different sources of signals carefully, our method beats state-of-the-art recommendation methods especially in user \emph{cold-start} settings. 
% We also merge the embeddings of the our model, to further adapt the \emph{coldness}. 
%Our model is 
%benchmarked against baselines on 
We evaluated our model on
four large, real-world datasets---Ciao, Foursquare, Epinions, and Flixster. 
%We performed comprehensive 
Our
experiments demonstrate that the model is capable of tacking different levels of \emph{cold-start} issues.
%JULIAN: Probably not significant enough for the conclusions
%, efficiency for inferring model parameters, as well as sensitivity to different hyperparameters.

%JULIAN: Sounds reasonable, but obvious enough that reviewers may question why you didn't simply try it.
%For further work, we are considering using a time-decaying scheme to further model the different amount of impact from activities that occurred long time ago within the social communities verses those happened just recently.
% Our current model learns the affinity between different users by inferring from the data.  actively select the `best' friends from one user's relations, and exclude those `strangers' from the training process; (2) we can set a time decay when one user refers to his/her friends' feedback to exclude the feedback which is too old. 
%Uncovering rating dimensions and modeling user-item interactions upon them are key to building a successful recommender system. In this paper, we proposed a sparse hierarchical embedding method, \emph{\model}, that simultaneously reveals globally-relevant and subtle visual dimensions efficiently. We evaluated \model~for personalized ranking tasks in the one-class setting and found it to significantly outperform state-of-the-art methods on real-world datasets. %in both \emph{warm-start} and \emph{cold-start} settings.

\section*{Appendix A}

Partial derivatives of $\widehat{x}_{u,i,l} - \widehat{x}_{u,j,l}$ with respect to our parameters are given by:

\begin{equation}
\begin{aligned}
\frac{\partial}{\partial \beta_i} &= 1;~~~\frac{\partial}{\partial \beta_j} = -1	\\
\frac{\partial}{\partial \gamma_i^I} &= \gamma_u^U;~~~\frac{\partial}{\partial \gamma_j^I} = -\gamma_u^U; ~~~\frac{\partial}{\partial \gamma_u^U} = \gamma_i^I - \gamma_j^I \\
\frac{\partial}{\partial \theta_i^I} &= \theta_l^I; ~~~ \frac{\partial}{\partial \theta_j^I} = -\theta_l^I;~~~\frac{\partial}{\partial \theta_l^I} = \theta_i^I - \theta_j^I\\
\frac{\partial}{\partial M_i} &= \frac{2}{|\mathcal{F}_u|^\alpha}\sum_{u'\in \mathcal{F}_u, i'}\sigma\big(\langle W_u, W_{u'}\rangle\big) \cdot M_{i'}\\
\frac{\partial}{\partial M_j} &= -\frac{2}{|\mathcal{F}_u|^\alpha}\sum_{u'\in \mathcal{F}_u, i'}\sigma\big(\langle W_u, W_{u'}\rangle\big) \cdot M_{i'}\\
\frac{\partial}{\partial M_{i'}} &= \frac{2}{|\mathcal{F}_u|^\alpha} \sigma\big(\langle W_u, W_{u'}\rangle\big) \cdot (M_i - M_j)\\
\frac{\partial}{\partial W_{u'}} &= \frac{2}{|\mathcal{F}_u|^\alpha} \sigma'\big(\langle W_u, W_{u'}\rangle \big) \cdot \langle M_i-M_j, M_{i'}\rangle  \cdot W_u \\
\frac{\partial}{\partial W_{u}} &= \frac{2}{|\mathcal{F}_u|^\alpha} \!\!\!\! \sum_{u'\in \mathcal{F}_u, i'} \!\!\!\! \sigma'\big(\langle W_u, W_{u'}\rangle\big) \cdot \langle M_i-M_j, M_{i'}\rangle \! \cdot\! W_{u'}
\end{aligned}
\label{eq:derivative}
\end{equation}
where $\sigma'(z)$ is the derivative of the sigmoid function, i.e., $\sigma(z) \cdot \sigma(-z)$. 
%JULIAN: I can't see a good reason to have this in addition to the derivatives. It seems to be just describing regular gradient ascent, there's nothing new about the algorithmic process
% Our complete learning procedure is presented in Algorithm~\ref{algo:algo}.
% \RestyleAlgo{boxruled}
% \begin{algorithm}[t] 
%  \SetKwInOut{Input}{Input}\SetKwInOut{Output}{Output}
%  \SetKwInOut{Initialization}{Initialization}
%  \SetKwInOut{Train}{Train}
%  \Input{The positive feedback pair set $P=\{(u, i)\}$, and the social network $G$}
%  \Output{The parameter set $\Theta$}
%  \Initialization{Initialize the parameters in $\Theta$\;}
%  \For{$u\leftarrow 1$ \KwTo $|\mathcal{U}|$}{
%    Split $u$'s items into training set, validation set and test set\;
%  }
%  \Train{}
%  \For{\#iterations}{
%  	\For{\#training samples}{
%       Uniformly sample a user $u$\;
%       Uniformly sample an item $i$ from $u$'s positive feedback\;
%       Uniformly sample an item $j$ where $j \neq i$\;
%       Obtain the last item $l$ from $i$\;
%       Obtain friends' last items $\{i'\}$ from the social network $G$ and the feedback history\;
%       Calculate $\frac{\partial (\widehat{x}_{u,i,l} - \widehat{x}_{u,j,l})}{\partial \Theta}$ according to Equation~\ref{eq:derivative}\;
%       Update $\Theta$ according to Equation~\ref{eq:update}\;
%     }
%  }

%  \caption{Stochastic gradient ascent of \model{}}
%  \label{algo:algo}
% \end{algorithm}

% \newpage
% \setlength{\bibsep}{0pt}
\bibliographystyle{named}
% \small
\bibliography{spmc}

\end{document}